\documentclass[12pt]{article}
\usepackage{graphics}
\usepackage{amsfonts}
\usepackage{amsmath}
\usepackage{amssymb}
\let\a=\alpha     \let\d=\delta 
     \let\th=\theta  
             \let\p=\pi    
    \let\f=\phi 
\let\ps=\psi   \let\o=\omega

  \let\del=\nabla

\def\\{\hfill\break} \let\==\equiv

\let\0=\noindent
\def\bra#1{{\langle#1|}}\def\ket#1{{|#1\rangle}}
\def\media#1{{\langle#1\rangle}}

\let\dpr=\partial

\def\qed{\hfill\raise1pt\hbox{\vrule height5pt width5pt depth0pt}}
      
\let\ra=\rightarrow 

\def\be{\begin{equation}}
\def\ee{\end{equation}}
\def\bea{\begin{eqnarray}}\def\eea{\end{eqnarray}}

\begin{document}
\markright{Quantum Interference...}
\title{``Quantum Interference with Slits" Revisited}

\author{Tony Rothman$^*$ and Stephen Boughn$^\dagger$  
\\[2mm]
{\small\it \thanks{trothman@princeton.edu}}~ \it Princeton University, \\
\it Princeton, NJ 08544 \\
{\small\it\thanks{sboughn@haverford.edu}}~ \it Haverford College, \\
\it Haverford PA, 09140}

\date{{\small   \LaTeX-ed \today}}

\maketitle

\begin{abstract}
Marcella has presented a straightforward technique employing the Dirac formalism to calculate single- and double-slit interference patterns.  He claims that no reference is made to classical optics or scattering theory and that his method therefore provides a purely quantum mechanical description of these experiments.  He also presents his calculation as if no approximations are  employed.  We show that he implicitly makes the same approximations found in classical treatments of interference and that no new physics has been introduced.  At the same time, some of the quantum mechanical arguments Marcella gives are, at best, misleading.

\vspace*{5mm} \noindent PACS: 3.65.--w, 03.65Ca, 3.65.TA, 01.70.+w
\\ Keywords: Quantum mechanics, two-slit experiment, interference, scattering theory

\end{abstract}

\baselineskip 8mm
In a 2002 paper Marcella\cite{Mar02} pointed out that quantum mechanics textbooks typically ``finesse" discussion of the famous double-slit interference experiment by resorting to analogies from classical wave optics in order to obtain the standard results.  He goes on to present a simple method of calculating a single- or double-slit interference pattern via the Dirac formalism and claims that this provides a quantum description of the experiment, moreover one that should be accessible to upper-level undergraduates.  It is  true that the usual textbook discussions of the two-slit experiment are riddled with ambiguities and omissions\cite{SR91}; and this is for the  reason that a consistent, quantum-mechanical treatment of the two-slit experiment does not, to our knowledge, exist.  Such a treatment would necessarily  introduce an interaction between the incident particle beam and slits that results in a probabilistic distribution of particles on the observation screen. The Hamiltonian governing this interaction would, in general, be quite complicated, and it is far from clear that even in principle it could resolve the ``measurement problem," the abrupt change of a deterministic system to a probabilistic one at the instant of measurement. One might therefore question whether Marcella's approach, which includes no explicit interactions---in particular no solution to a wave equation---can claim to give an adequate accounting of the two-slit experiment.  Furthermore, at first glance he makes makes virtually no assumptions whatsoever and makes no reference to any approximations, leaving the impression that his result is exact. We now show that he has merely carried out the same classical interference calculation that students encounter in first- or second-year physics, and with similar assumptions.

\begin{figure}[htb]
\vbox{\hfil\scalebox{.5}
{\includegraphics{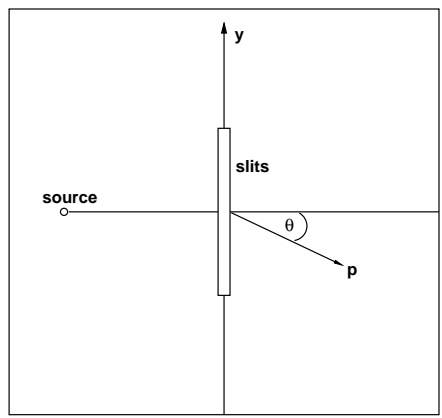}}\hfil}
{\caption{\footnotesize{Marcella's setup for the one-and two-slit experiment.  After passing through the slit(s), the particle is assumed to be scattered at an angle $\th$ with a momentum $p$ and a
y-momentum of $p_y = p\sin\th$.   Note that
$\th$ is measured from the centerline.}}}
\end{figure}

We focus on the single-slit experiment; the double-slit goes through in exactly the same manner.   The following is essentially Marcella's analysis, with his experimental arrangement shown in figure 1.  Note that the angle $\th$ is drawn from the center of the single slit or the midpoint between the two slits. If the system, in this case a particle, is initially in a state $\ket\ps$, then according to the Born postulate, the probability $P$ that the  particle is scattered with y-momentum $p_y = p\sin\th$ is  given by
\be
P(p_y) = |\media{p_y|\ps}|^2, \label{prob}
\ee
where the bracket represents the overlap integral between the initial and final states.  In the position representation, the momentum eigenfunction for a free particle with momentum $p_y$ is
\be
\f \equiv  \media{y|p_y} = \frac1{\sqrt 2\p} e^{\frac{ip_yy}{\hbar}}.
\ee
The scattering amplitude  $\media{p_y|\ps}$ can be calculated by inserting a complete basis of position states between $\bra{p_y}$ and $\ket\ps$  and integrating:
\be
\media{p_y|\ps} = \int_{-\infty}^\infty \media{p_y|y}\media{y|\ps} dy
= \frac1{\sqrt 2\p}\int_{-\infty}^\infty e^{-\frac{ip_yy}{\hbar}}\ps(y) dy, \label{scatamp}
\ee
where in the Dirac formalism the position wave function $\ps(y) \equiv \media{y|\ps}$.  To calculate $P(p_y)$ from this expression we need to know $\ps(y)$ at the position of the slit(s).  Taking the slit width to be $a$, Marcella makes the most natural assumptions:
\be
\ps(y) = \media{y|\ps} = \left\{
  \begin{array}{ll}
        1/\sqrt{ a},     & -a/2 \leq y  \leq a/2 \\
        0,     &  \mbox{elsewhere.}
  \end{array}
  \right. \label{intensity}
\ee
The first statement is merely  that the intensity is assumed to be uniform over the slit, or that there is an equal likelihood of finding the  particle anywhere within it.  In other words, we take $|\ps(y)|^2dy = C dy$ for constant $C$.  Integrating over the slit gives the total probability of finding the particle $P = C a \equiv 1$.  Therefore $C = 1/a$ and  $\ps(y) = 1/\sqrt a$.

 We now simply plug this expression for $\ps$ into Eq. (\ref{scatamp}) to calculate the scattering amplitude:
 \bea
 \media{p_y|\ps} &=& \frac1{\sqrt{ 2\p a}}\int_{-a/2}^{a/2} e^{-\frac{ip_yy}{\hbar}}dy\nonumber\\
                 &=& \frac{i\hbar}{p_y\sqrt{ 2\p a}}(e^{-\frac{ip_y a}{2\hbar}}- e^{\frac{ip_y a}{2\hbar}})\nonumber\\
                 &=& \frac{2\hbar}{p_y\sqrt{ 2\p a}}\sin\left(\frac{ap_y}{2\hbar}\right).\label{scatint}
 \eea
With the definition $\a = a p_y/{2\hbar}$, Eq. (\ref{prob}) gives for the probability of detecting a particle at angle $\th$
\be
 P(y)= \frac{a}{2\p}\left(\frac{\sin\a}{\a}\right)^2, \label{standresult}
 \ee
 the standard result for single-slit diffraction.\\

Upon being presented with this procedure, one's first reaction might well be, ``Where's the physics?"  The Dirac formalism is exactly that, a formalism, and does not in and of itself add any  physics to the problem.  We see that the above technique includes no wave equation to be satisfied and no interaction---in other words, no dynamics. Has a miracle occurred?

In classical optics one generally treats the scalar diffraction problem by finding a solution to the Helmholtz equation:
\be
(\del^2 + k^2)\ps = 0,
\ee
of which the time-independent Schr\"odinger equation is the special case  $k^2  = 2m(E-V)/\hbar^2 $, for constant $E$ and $V$.  The Helmholtz equation cannot be solved exactly except in nearly trivial circumstances, and so one must resort to numerous approximations.  The usual ones for scalar diffraction theory are:

1) The ``freshman physics approximation,"  or basic Huygens construction, which assumes that each point within the slit acts as a source of spherical waves $\psi(r) = e^{ikr}/r$ radiating only in the forward direction.  Note that this wave function is a solution to the source-free Helmholtz equation;

\begin{figure}[htb]
\vbox{\hfil\scalebox{.7}
{\includegraphics{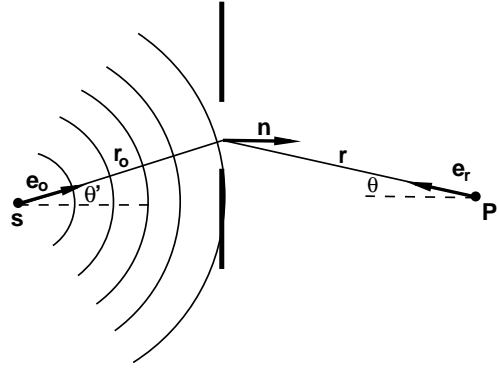}}\hfil}
{\caption{\footnotesize{A source $s$ emits a spherical wavefront whose unit normal ${\bf e}_o$ is at an angle $\th'$ with respect to the x-axis.  At the aperture the wavefront is diffracted to an observation point $P$.  The normal to the aperture is $\bf n$ and the unit vector from $P$ to a point on the aperture is ${\bf e}_r$. }}}
\end{figure}

2) More sophisticated treatments include an ``obliquity factor" in the Huygens construction; the solution is assumed to be of the form $\ps \sim f(\th, \th')e^{ikr}/r$, where $\th$  and $\th'$ are as shown in figure 2.  To find the obliquity factor, one generally employs the Kirchoff approximation \cite{EH69,Mar72,Jack75}, which in the case of an aperture in a plane screen assumes that a) the  wave function $\ps$ and $\dpr\ps/\dpr n \equiv \del\ps\cdot\bf n$, for $\bf n$ the unit normal to the aperture in the direction of propagation, are zero everywhere except in the aperture; b) in the aperture $\ps$ and $\dpr\ps/\dpr n$ take on the value of the incident wave.  Any text on electromagnetism or optics, including those just cited, shows that with the Kirchoff approximation the value of $\ps$ at the observation point $P$ is given by the integral
\be
\ps(P) = \frac{iAk}{4\pi}\int_a \frac{e^{ik(r+r_o)}}{rr_o}({\bf e}_r - {\bf e}_o)\cdot {\bf n}\, da,
\ee
where $A$ is a constant and the integration is over the aperture.  As one sees from figure 2, the Kirchoff integral can be written
\be
\ps(P) = -\frac{iAk}{4\pi}\int_a \frac{e^{ik(r+r_o)}}{rr_o}(\cos\th + \cos\th')\, da, \label{kirch2}
\ee
where the obliquity factor is $f(\th, \th') = (1/2)(\cos\th + \cos\th')$.

It is well known that specifying both $\ps$ and $\dpr\ps/\dpr n $ overdetermines the system and makes the Kirchoff approximation inconsistent; nevertheless it generally gives results that are in good agreement with experiment for small angles in the short wavelength limit.  One can remove the inconsistencies by choosing one boundary condition or the other; that is, by specifying either $\ps$ within the aperture  or $\dpr\ps/\dpr n$ but not both.  Thus we have the other two approximations:

3) Dirichlet  approximation, in which $\ps$ is taken to be the indicident wave within the aperture.  For the case of a plane aperture, the obliquity factor becomes $f = \cos(\th)$;

4) Neumann approximation, in which $\dpr\ps/\dpr n$  is specified within the aperture.  Here for a plane screen the obliquity factor becomes $f = \cos(\th')$

None of these four methods lead to exact solutions of the Helmholtz equation and all differ in the limit of large scattering angles and long wavelengths.  However, all four agree in the standard slit diffraction regime of normal incidence, small scattering angles and short wavelengths.  In this limit, with the additional Fraunhofer approximation that  the distance to the screen is large compared to the aperture width, the Kirchoff integral for an infinite slit of width $a$ in the y-direction becomes
\be
\ps(P) = C\int_{-a/2}^{a/2}e^{-ik\sin\th y} dy, \label{fraun}
\ee
where now $\th$ is the fixed angle between $P$ and the center of the slit.\\

The question is, Has Marcella made implicit use of any of this formalism?  At first glance it appears not; he does not employ the   Schr\"odinger equation and makes no reference to any approximations.  He is purely concerned with what happens to the particle \emph{at the slit} and makes no statement about what takes place at the distant observation screen.  This is similar to a ``hidden-variable" point of view, that the particle is emitted from the screen with a definite momentum before a measurement takes place.  It is just this sort of interpretative statement that leads to analyses that violate Bell's inequalities and is certainly not the sort of concept that should be implanted in the mind of a student of quantum mechanics. If, however, the discussion is to take place within the framework of standard quantum theory, we have the right to demand that the Schr\"odinger equation be satisfied, as well as quantum mechanics' standard postulates. Thus, Marcella must be implicitly solving the Helmholtz equation in some limit.

We can uncover this limit by a close examination of Marcella's few assumptions.   Note first that the assumption of equal intensity across the slit made in Eq. (\ref{intensity}) is equivalent to the statement that the incoming wavefront is a plane wave.  The normal vector ${\bf n}_o$ of a spherical wavefront impinging on a plane aperture would vary in angle with respect to the screen normal $\bf n$ across the aperture; hence the relationship of the area elements and intensity could not be constant on the screen.  This does not necessarily imply normal incidence; however, since Marcella does not specify phases, it might as well be.  That is, the intensity could either be $|\ps_o|^2$ from waves normal to the aperture, or $|\ps\cos(\th')|^2 = |\ps_o|^2$ for non-normal wave function $\ps$.  But since  $\th'$ must be constant, it can be absorbed into the definition of $\ps$.  Furthermore, Marcella has assumed that $\ps = 0$ outside the aperture, but has not said anything about $\dpr\ps/\dpr n$.  Thus he has implicitly assumed a Dirichlet boundary condition, and so the obliquity factor should be $\cos(\th)$, which has been set to one, i.e., the paraxial approximation has been used.

At the same time, we notice that in the integration of Eq. (\ref{scatint}), he has held $p_y = p\sin(\th)$ constant.  He justifies this by saying that he is calculating the scattering amplitude of a particle being emitted \emph{at the slit} with a fixed angle $\th$.  As mentioned, he makes no statement about what takes place at the distant observation point.  In standard quantum mechanics, however, one would  assume that the wave function could be emitted at different angles from the top and bottom slit or integrate over $\th$.  That Marcella uses a single angle (figure 1) drawn from the center of the slit, or from the midpoint of the double-slit configuration is permissible only in the Fraunhofer limit.  Thus, at best, he is performing the integral in Eq. (\ref{fraun})

However, one can see that his assumptions that $\ps = 1/\sqrt a$ and use of the plane-wave eigenstates in Eq. (\ref{scatint}) is not consistent.  All that is being done in Eq.(\ref{scatint}) is to take the Fourier transform of $\ps(y) = 1/\sqrt a$.  If we transform from position to momentum space and back again, we should end up where we started: $\ps'(y) = \ps(y)= 1/\sqrt a$.  Thus,
\be
\ps'(y) = \frac1{\sqrt{2\pi}}\int_{-k_m}^{k_m} \f(k)e^{iky} dk,
\ee
where we have let the limits of integration be finite and where
\be
\f(k) \frac1{\sqrt a} \int_{-a/2}^{a/2}e^{-iky'}dy'.
\ee
Hence,
\bea
\ps'(y) &=& \frac1{{2\pi}\sqrt a}\int_{-k_m}^{k_m} \int_{-a/2}^{a/2}e^{ik(y-y')}dy' dk,\nonumber\\
&=& \frac1{{2\pi}\sqrt a}\int_{-a/2}^{a/2} dy'\int_{-k_m}^{k_m}  e^{ik(y-y')} dk.
\eea
If $k_m \ra \infty$, then the last factor becomes $2\p\d(y-y')$ and the result $\ps'(y) = 1/\sqrt a$ as required.  However, in Marcella's problem $p_y = \hbar k_y = p\sin(\th)$ and there is a maximum value of $k$.  In that case
\be
\ps'(y) = \frac1{\pi \sqrt a}\int_{-a/2}^{a/2} \frac{\sin(k_m(y-y'))}{(y-y')}dy',
\ee
which is nonanalytic and must be approximated for a given value of $y$.  For $k_m(y-y') << 1$ one gets Marcella's result only when, interestingly, $\lambda = 2a$, but in that case the assumption that $k_m(y-y') << 1$ cannot always be met.

\begin{figure}[htb]
\vbox{\hfil\scalebox{.7}
{\includegraphics{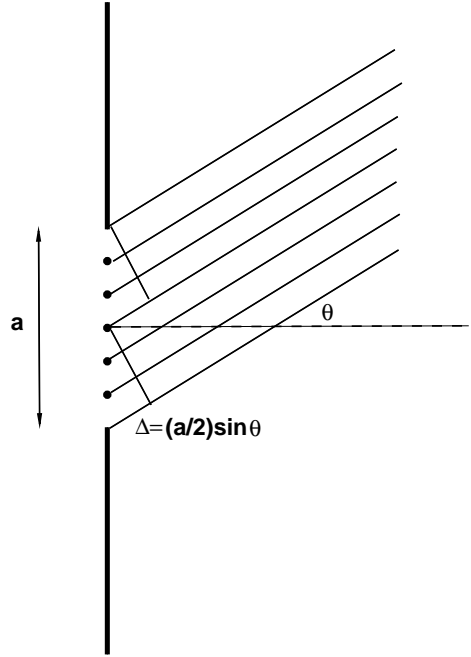}}\hfil}
{\caption{\footnotesize{Consider a slit of width $a$ and divide it into $2N$ sources of Huygens wavelets, $N$ above the central source and $N$ below.  For parallel rays the phase shift with respect to the central source will be $\pm k(a/2)\sin\th(n/N)$ for the n-th source.  The phase shift at the end of the slit is $\Delta =\pm k(a/2)\sin\th $ }}}
\end{figure}

In fact Marcella's approximations are no better than the freshman physics approach, which ignores all boundary conditions.  Consider figure 3, in which we divide up a slit of width $a$ into $2N$ Huygens sources, $N$ above the central source and $N$ below.  Assume that the screen is at an infinite distance so that the rays are parallel.  At the observation point $P$ on the screen the wave function is merely the addition of all the Huygens wavelets. Thus
\be
\ps(P) = \psi_o(P)\sum_{n = -N}^N e^{i(\o t + \d_n)}, \label{Huygens}
\ee
where the $n^{th}$ phase shift with respect to the central source is $\d_n = k(a/2)\sin(\th)n/N$.  Write $\d_n = \a_n sin\th$ with $\a_n \equiv k(a/2)n/N$.  Notice that $\a_n$ plays the role of $k$ or momentum, while $\th$ is the position variable.  The spacing between adjacent $n$'s is $\Delta\a = (ka/2N) = constant$.   Multiplying and dividing Eq. (\ref{Huygens}) by $\Delta\a$ gives
\be
\ps(P) = \frac{\psi_o(P)}{\Delta\a} \sum_{n = -N}^N e^{i\a_n\sin\th}\Delta\a,
\ee
where we have absorbed the maximum value (say) of the rapidly varying function $e^{i\o t}$ into the definition of $\ps_o$.  Absorb the first $\Delta \a$ into the definition of $\ps_o$ as well.  Then, in the limit of large $N$,
\bea
\ps(P) &\ra & \psi_o(P) \int_{-ka/2}^{ka/2} e^{i\a\sin\th} d\a,\nonumber\\
       &=& \ps_o(P) \frac{2\sin((ka/2)\sin\th)}{\sin\th}
\eea
which is exactly the standard answer.  We thus see that the freshman physics construction of adding together Huygens wavelets is really a Fourier transform, which is exactly what Marcella has introduced by chopping his wave function at the edge of the slits. Although he is going from position to momentum space, by writing the result Eq. (\ref{standresult}) in terms of $\th$, we have the same position-space result just obtained.

In sum, Marcella does make the valid point that quantum interference should be treated as a quantum phenomenon and quantum texts ought not immediately redirect the discussion to classical wave optics.  But a more reasonable way to do this would be to simply show that the Schr\"odinger equation reduces to the Helmholtz equation, thus reducing the problem to one of  classical scalar scattering with its concomitant approximations.  This would also provide the opportunity of discussing relevant boundary conditions and to point out the difficulty of specifying them precisely in both the quantum and electromagnetic cases.  As it stands, while Marcella's procedure is useful in giving students practice with the Dirac formalism,  it has introduced no quantum physics into the problem other than setting $p = \hbar k$, and has implicitly made all the assumptions that show this is indeed a problem of classical optics.  That his result is  the same as the one obtained by the simplest Huygens construction is merely a reflection of the fact that he has implicitly made the lowest-order approximations, where all methods converge to the same result. \\

{\small

 \end{document}